\PassOptionsToPackage{unicode}{hyperref}
\PassOptionsToPackage{hyphens}{url}
\PassOptionsToPackage{dvipsnames,svgnames,x11names}{xcolor}
\documentclass[
  11pt,
]{article}
\usepackage{amsmath,amssymb}
\usepackage{setspace}
\usepackage{iftex}
\ifPDFTeX
  \usepackage[T1]{fontenc}
  \usepackage[utf8]{inputenc}
  \usepackage{textcomp} 
\else 
  \usepackage{unicode-math} 
  \defaultfontfeatures{Scale=MatchLowercase}
  \defaultfontfeatures[\rmfamily]{Ligatures=TeX,Scale=1}
\fi
\usepackage{lmodern}
\ifPDFTeX\else
\fi
\IfFileExists{upquote.sty}{\usepackage{upquote}}{}
\IfFileExists{microtype.sty}{
  \usepackage[]{microtype}
  \UseMicrotypeSet[protrusion]{basicmath} 
}{}
\makeatletter
\@ifundefined{KOMAClassName}{
  \IfFileExists{parskip.sty}{%
    \usepackage{parskip}
  }{
    \setlength{\parindent}{0pt}
    \setlength{\parskip}{6pt plus 2pt minus 1pt}}
}{
  \KOMAoptions{parskip=half}}
\makeatother
\usepackage{xcolor}
\usepackage[margin=1in]{geometry}
\usepackage{color}
\usepackage{fancyvrb}

\DefineVerbatimEnvironment{Highlighting}{Verbatim}{commandchars=\\\{\}}
\newenvironment{Shaded}{}{}

\newcommand{\DataTypeTok}[1]{\textcolor[rgb]{0.56,0.13,0.00}{#1}}

\newcommand{\ErrorTok}[1]{\textcolor[rgb]{1.00,0.00,0.00}{\textbf{#1}}}

\newcommand{\FunctionTok}[1]{\textcolor[rgb]{0.02,0.16,0.49}{#1}}

\newcommand{\OtherTok}[1]{\textcolor[rgb]{0.00,0.44,0.13}{#1}}

\newcommand{\StringTok}[1]{\textcolor[rgb]{0.25,0.44,0.63}{#1}}

\usepackage{longtable,booktabs,array}
\usepackage{calc} 
\usepackage{etoolbox}
\makeatletter
\patchcmd\longtable{\par}{\if@noskipsec\mbox{}\fi\par}{}{}
\makeatother
\IfFileExists{footnotehyper.sty}{\usepackage{footnotehyper}}{\usepackage{footnote}}
\makesavenoteenv{longtable}
\providecommand{\tightlist}{%
  \setlength{\itemsep}{0pt}\setlength{\parskip}{0pt}}
\usepackage{microtype}
\usepackage{booktabs}
\usepackage{xurl}
\ifLuaTeX
  \usepackage{selnolig}  
\fi
\usepackage[]{natbib}
\IfFileExists{bookmark.sty}{\usepackage{bookmark}}{\usepackage{hyperref}}
\IfFileExists{xurl.sty}{\usepackage{xurl}}{} 
\hypersetup{
  pdftitle={Notarized Agents: Receiver-Attested Confidential Receipts for AI Agent Actions},
  colorlinks=true,
  linkcolor={Maroon},
  filecolor={Maroon},
  citecolor={Blue},
  urlcolor={Blue},
  pdfcreator={LaTeX via pandoc}}

\title{Notarized Agents: Receiver-Attested Confidential Receipts for AI
Agent Actions}
\author{Juan Figuera\\
Independent Researcher, Sello Project\\
\texttt{juan@figuera.co}}
\date{Draft of May 30, 2026}

\begin{document}
\maketitle
\begin{abstract}
Current AI agent observability is structurally compromised: the entity
producing the activity log is the same entity whose activity is being
logged. A compromised or buggy agent can omit, alter, or fabricate its
own traces, and the operator running the agent has no independent way to
detect tampering. We propose a class of protocols that resolves this by
inverting the trust boundary: the service that receives an agent's call
signs a receipt of what it observed using its own key, encrypts the
receipt to the agent's owner, and publishes it to a public transparency
log. The owner reconstructs a tamper-evident trail without trusting the
agent or its operator. We instantiate the class as Sello, a protocol
combining four properties absent in any current system: (P1)
receiver-side signing, (P2) HPKE encryption to an owner public key bound
to the authorization token via JWS, (P3) publication to a
witness-cosigned Merkle log, and (P4) owner-side discovery by token
reference. We describe the protocol, analyze its security under an
adversary that controls the agent and its operator, present
microbenchmarks of the cryptographic operations, and situate Sello among
adjacent receipt-protocol work (Signet, AgentROA, Agent Passport System,
draft-farley-acta, SCITT). We discuss known limitations including the
suppression attack, service collusion, and the adoption-incentive
problem.
\end{abstract}

\setstretch{1.15}
\hypertarget{introduction}{%
\subsection{1 Introduction}\label{introduction}}

Agents acting on behalf of humans need to be auditable by those humans.
The agent's owner needs to know not just what the agent claims it did,
but what it actually did. This requirement is foundational to deploying
agents in any context where the consequences of agent action are
non-trivial: financial transactions, healthcare data access, code
modification, communication on the principal's behalf.

The dominant approach to agent observability today is to instrument the
agent itself, the agent's runtime, or a proxy in the operator's
infrastructure. Platforms such as Langfuse, LangSmith, Helicone,
Braintrust, and AgentOps collect traces from inside the agent's process
and aggregate them for the owner's inspection. This is the model that
has shipped to production at scale, and the model that every commercial
agent observability product currently in the market follows.

This model fails the threat model that agent deployment actually faces.
The agent is the actor; the agent is also the source of truth about its
own behavior. When the agent is compromised (by prompt injection,
context poisoning, tool-call hijacking, or simply buggy code), the
traces it produces about its own actions cannot be trusted, because the
same code path that misbehaves is the code path that writes the trace.
When the operator running the agent has misaligned incentives, the
traces can be silently rewritten before they reach the owner. The actor
is not a reliable narrator of its own actions, and we have known this
for the entire history of accounting, law, journalism, and science.
Recent work on AI auditing makes a related argument, that black-box
query access alone is insufficient for rigorous audits and that auditors
need richer forms of access to assess a system's behavior
\citep{casper-blackbox}; receiver-attested receipts are one
cryptographic answer to that broader question.

We can sharpen the requirement concretely. A record of an agent action
should be:

\begin{itemize}
\tightlist
\item
  \textbf{unforgeable} by the agent or its operator, even given full
  control of the agent's runtime;
\item
  \textbf{independently verifiable} by the owner, using only public
  infrastructure and owner-held keys;
\item
  \textbf{confidential} from the operators of any public infrastructure
  used to store or transmit it;
\item
  \textbf{attributable} to a party that was actually present when the
  action occurred and is structurally separate from the agent;
\item
  \textbf{retrievable} by the owner after the interaction is complete,
  even when the agent is unavailable or compromised.
\end{itemize}

No system we surveyed delivers all five.

This paper applies a different architectural principle, one with deep
precedent outside computing. In every legal, scientific, and commercial
system where the truthfulness of a record matters, the record is written
by a party who was present at the event but is independent of the
principal. Mesopotamian commercial contracts circa 3000 BCE were
impressed into clay tablets witnessed by named third parties whose seals
were unique to each. Roman notaries inherited the practice. Medieval
scribes, customs officers, court reporters, lab technicians, and modern
notarization systems all rest on the same idea. When the recording party
is structurally separate from the recorded party, the record carries
weight.

Apple's \emph{Find My} network applies this principle in modern
infrastructure. AirTags do not phone home; they cannot, because they are
passive Bluetooth devices with no internet connection. Instead, every
iPhone in the world that walks past an AirTag submits an end-to-end
encrypted ``I saw tag X near me'' report that only the tag's owner can
decrypt. The tracked object is dumb. The witnesses, independent iPhones,
none controlled by the tag's owner, are what produce the verifiable
record \citep{Heinrich2021}.

We port this architectural inversion to AI agent observability. The
agent is the tracked object. Every service the agent talks to (MCP
server, API, tool endpoint, A2A peer) is a witness. Each witness signs a
receipt of the interaction it observed using its own cryptographic key,
encrypts the receipt to the agent owner's public key, and publishes the
encrypted receipt to a public transparency log. The owner reconstructs
the agent's activity by querying the log, verifying signatures against a
registry of service identities, and decrypting receipts locally. The
agent never holds any of the signing keys involved. The operator never
holds the owner's encryption key. A compromised agent cannot fake its
trail because the trail is signed by parties it does not control.

We call this class of protocols \emph{notarization for AI agents}.
Members of the class can differ in their authorization-token format,
their choice of transparency log, and their service-identity system,
while sharing the four defining properties. We instantiate the class as
\textbf{Sello}, a protocol that combines four properties absent in any
system we surveyed: receiver-side signing by the called service (P1),
HPKE encryption to an owner public key cryptographically bound to the
authorization token (P2), publication to a witness-cosigned Merkle log
(P3), and owner-side discovery by token reference (P4). The combination
is the contribution.

\hypertarget{contributions}{%
\subsubsection{Contributions}\label{contributions}}

This paper makes four contributions:

\begin{enumerate}
\def\labelenumi{\arabic{enumi}.}
\item
  We identify the trust-boundary inversion as a structural fix for AI
  agent observability and articulate the four cryptographic properties
  (P1-P4) that distinguish notarized-agent protocols from existing
  approaches.
\item
  We specify Sello, a concrete instantiation built on standard
  primitives (COSE\_Sign1, HPKE, Ed25519, Merkle transparency logs) and
  define a JWS token profile that cryptographically binds the owner's
  HPKE encryption key to the authorization token presented by the agent.
\item
  We propose using the transparency log's integrated time as the binding
  clock for key revocation decisions, separating it from the
  service-asserted timestamp in the receipt body. This addresses a class
  of attacks where a compromised service backdates receipts to predate
  its key compromise.
\item
  We provide a property-by-property comparison of Sello against existing
  receipt-protocol work (Signet, AgentROA, Agent Passport System,
  draft-farley-acta, agentreceipts.ai, Pipelock, Attested Intelligence,
  and the SCITT working group), showing that no current system combines
  all four properties.
\end{enumerate}

We also honestly state what Sello does not solve: the suppression attack
(the agent simply not calling services produces no receipts), service
collusion, and the adoption-incentive problem (services must choose to
emit receipts; today, none have reason to).

The remainder of the paper is organized as follows. Section 2 covers
background and related work. Section 3 specifies the threat model and
design goals. Section 4 specifies the Sello protocol. Section 5 analyzes
its security properties under the threat model. Section 6 distinguishes
per-receipt verifiability from set-completeness and discusses retrieval
guarantees. Section 7 presents microbenchmarks. Section 8 discusses
limitations, the adoption-incentive problem, and future extensions.
Section 9 concludes.

\begin{center}\rule{0.5\linewidth}{0.5pt}\end{center}

\hypertarget{background-and-related-work}{%
\subsection{2 Background and Related
Work}\label{background-and-related-work}}

We briefly review the cryptographic primitives Sello composes, then
survey existing work on cryptographic receipts for AI agent actions.

\hypertarget{cryptographic-primitives}{%
\subsubsection{2.1 Cryptographic
primitives}\label{cryptographic-primitives}}

\textbf{COSE\_Sign1} \citep{RFC9052} is the CBOR Object Signing and
Encryption single-signer signature structure. It is the binary
equivalent of JWS for CBOR-encoded payloads. We use COSE\_Sign1 as the
signing envelope for Sello receipts because it is the format used by the
IETF SCITT working group for transparency receipts
\citep{scitt-architecture}, giving Sello a natural alignment path for
standardization.

\textbf{HPKE (Hybrid Public Key Encryption)} \citep{RFC9180} is a
standardized public-key encryption scheme that combines a Key
Encapsulation Mechanism (KEM), a Key Derivation Function (KDF), and an
Authenticated Encryption with Associated Data (AEAD) primitive. Sello
uses HPKE in single-shot mode with X25519 for the KEM, HKDF-SHA256 for
the KDF, and ChaCha20-Poly1305 for the AEAD. HPKE is used in TLS
Encrypted Client Hello and MLS, giving it deep production validation.

\textbf{JWS (JSON Web Signature)} \citep{RFC7515} is the JSON
serialization for signed payloads. Sello defines a JWS-based
authorization token profile in which the owner's HPKE public key is
carried as a claim. We use compact serialization. Other token formats
(UCAN, biscuits, macaroons) would require their own analogous profiles
in future versions.

\textbf{Transparency logs} as we use them are append-only Merkle trees
with witness cosigning, descending from Certificate Transparency
\citep{RFC6962, RFC9162}. The witness cosigning idea originates with
Syta et al.'s CoSi protocol \citep{Syta2016}, which formalized the
principle that an authoritative statement should be validated and
publicly logged by a diverse group of witnesses before any client
accepts it. Modern instances include Sigstore Rekor
\citep{sigstore-rekor}, which Sello uses as its v0.1 reference log.

\hypertarget{receipt-protocols-for-ai-agent-actions}{%
\subsubsection{2.2 Receipt protocols for AI agent
actions}\label{receipt-protocols-for-ai-agent-actions}}

The ``receipts'' vocabulary for agent actions has emerged across several
projects in 2024-2026, each making different architectural choices. We
summarize each and identify which of Sello's four properties they have
and which they lack. Section 7.4 contains the full comparison table.

\textbf{Agent Receipts} \citep{agentreceipts} is an open specification
maintained by Otto Jongerius defining cryptographically signed agent
action receipts using Ed25519 and W3C Verifiable Credentials. The author
frames it as ``C2PA Content Credentials for agent actions.'' The signer
in Agent Receipts is the agent platform itself, which places the signing
key on the operator's side of the trust boundary.

\textbf{Signet} \citep{signet} is an MCP-focused middleware project
maintained by William Hou. v0.4 introduced bilateral co-signing in which
the MCP server holds its own Ed25519 key and signs response receipts.
v0.10 introduced envelope encryption using XChaCha20-Poly1305 with a key
derived from the signing identity. Signet is the closest existing system
to Sello's design, but the encryption key is not separated from the
signing identity, there is no public transparency log integration, and
the trust bundle is currently a hand-edited JSON file.

\textbf{Pipelock} \citep{pipelock} ships an out-of-process mediator
signer that runs as a sidecar to the agent's runtime. The architecture
explicitly distinguishes between in-process signing (the weakest trust
model), operator-deployed mediator signing (Pipelock's current shipping
mode), and independent third-party witness signing (listed as roadmap).
Pipelock has not shipped the third-party variant.

\textbf{Agent Passport System (APS)} \citep{aps} by Tymofii Pidlisnyi
defines Wave 1 accountability primitives consisting of four receipt
types: ActionReceipt, AuthorityBoundaryReceipt, CustodyReceipt, and
ContestabilityReceipt. ActionReceipt is signed by the executing agent
rather than the receiving service, which is a deliberate design choice
that addresses a different threat model than ours.

\textbf{draft-farley-acta-signed-receipts} \citep{farley-acta} is an
IETF Internet-Draft by Tom Farley at ScopeBlind defining a portable
signed receipt format for machine-to-machine access control decisions.
The reference implementation (protect-mcp) is a stdio proxy between
agent and MCP server, signing decision receipts on behalf of whoever
operates the proxy. Receipts use deterministic JSON canonicalization
with Ed25519. The draft does not specify encryption to a recipient and
does not integrate with a transparency log.

\textbf{draft-nivalto-agentroa-route-authorization}
\citep{nivalto-agentroa} is an IETF Internet-Draft by Joseph Michalak at
Nivalto defining Agent Enforcement Receipts (AERs) intended for
submission to SCITT-compatible transparency logs. The signer is the
egress border gateway operating under the operator's trust domain.
AgentROA has the SCITT-log piece of Sello's design but signs at
operator-controlled infrastructure rather than at the receiving service.

\textbf{Attested Intelligence} \citep{attestedintelligence} ships an MCP
governance proxy (\texttt{@attested-intelligence/aga-mcp-server}) that
emits Ed25519-signed receipts for every MCP tool call, hash-linked into
a tamper-evident continuity chain, packaged as offline-verifiable
evidence bundles. The company holds USPTO Patent Application 19/433,835
(filed December 28, 2025) on the ``Attested Governance Artifacts'' (AGA)
architecture, which uses standard cryptography (Ed25519, ML-DSA-65,
SHA-256, Merkle trees) and explicitly does not require Trusted Execution
Environments. The Portal that signs receipts sits in the operator's
trust boundary; the agent holds no keys. Receipts are distributed
point-to-point as evidence bundles rather than published to a
transparency log, and are not encrypted to an owner public key. Target
markets include defense, SCADA/ICS, regulated industries (finance,
healthcare), and enterprise agentic AI governance. The three independent
patent claims cover the Portal runtime with sealed policy and drift
detection, a substitution-based privacy mechanism with claims taxonomy,
and a continuity chain with payload-excluded leaf hashes; Sello uses
receiver-attested signing, HPKE encryption to owner, and a
payload-inclusive transparency log, none of which read on these claims.

\textbf{SCITT} (Supply Chain Integrity, Transparency, and Trust) is an
active IETF working group \citep{scitt-architecture, scitt-scrapi}
standardizing the COSE\_Sign1 transparency receipt framework. No current
SCITT WG draft applies the framework to AI agent tool calls; existing
profiles cover supply-chain artifacts (e.g.~CCF \citep{scitt-ccf}) and
one individual submission on financial trading audit
(draft-kamimura-scitt-vcp \citep{kamimura-scitt}).

\hypertarget{adjacent-academic-work}{%
\subsubsection{2.3 Adjacent academic
work}\label{adjacent-academic-work}}

Three recent papers are intellectually adjacent to Sello but make
different architectural choices.

\textbf{``Right to History: A Sovereignty Kernel for Verifiable AI Agent
Execution''} by Jing Zhang \citep{zhang-righttohistory} proposes a
personal-hardware sovereignty kernel (PunkGo) that produces
tamper-evident records of agent actions using an RFC 6962 Merkle tree
audit log local to the owner's machine. The architectural difference is
fundamental: Right to History assumes the owner has hardware-level
control of the agent's execution environment and can therefore observe
every action at the kernel level. Sello assumes the opposite: agents run
on hosted infrastructure outside the owner's control, and the owner
relies on the services the agent touches to attest to what they
observed.

\textbf{``Verifiability-First Agents''} by Abhivansh Gupta
\citep{gupta-verifiabilityfirst} proposes embedded Audit Agents that
continuously verify intent versus behavior. The verification is
structural rather than cryptographic: a second agent reads the first
agent's plans and outputs and checks for divergence. This complements
Sello (Audit Agents could consume Sello receipts as their ground truth)
but does not solve the underlying attestation problem.

\textbf{``Tool Receipts, Not Zero-Knowledge Proofs''} by Abhinaba Basu
\citep{basu-toolreceipts} proposes HMAC-signed receipts generated by the
agent's runtime to detect hallucinations. The architectural difference
is that the runtime is part of the agent's trust boundary, so the
receipts cannot defend against a compromised operator.

\begin{center}\rule{0.5\linewidth}{0.5pt}\end{center}

\hypertarget{threat-model-and-design-goals}{%
\subsection{3 Threat Model and Design
Goals}\label{threat-model-and-design-goals}}

\hypertarget{system-model}{%
\subsubsection{3.1 System model}\label{system-model}}

A Sello deployment has four actors:

\begin{itemize}
\tightlist
\item
  An \textbf{owner} who deploys an agent and holds a long-term HPKE key
  pair. The public key is bound to authorization tokens issued for the
  agent's use.
\item
  An \textbf{agent}, autonomous software acting on the owner's behalf,
  which presents authorization tokens when calling services.
\item
  A set of \textbf{services} the agent calls. Each service holds its own
  long-term Ed25519 signing key, independent of the owner and the
  operator.
\item
  A set of public \textbf{transparency logs} that accept signed entries,
  return Merkle inclusion proofs, and are cosigned by independent
  witnesses.
\end{itemize}

\hypertarget{adversary-model}{%
\subsubsection{3.2 Adversary model}\label{adversary-model}}

We assume a powerful adversary who may:

\begin{itemize}
\tightlist
\item
  Compromise the agent and arbitrarily edit its local logs, traces, and
  outputs.
\item
  Control the agent's operator infrastructure, including the runtime,
  the network path, and any operator-side telemetry or audit systems.
\item
  Read the public transparency log in full.
\item
  Submit arbitrary bytes to the log, including bytes that purport to be
  Sello receipts but cannot pass verification without possession of a
  legitimate service's signing key.
\end{itemize}

We assume the adversary does NOT possess:

\begin{itemize}
\tightlist
\item
  The Ed25519 signing keys held by legitimate services.
\item
  The HPKE private key of the agent owner.
\end{itemize}

These assumptions are non-trivial but standard. Service key compromise
is a known attack class, and the system must degrade gracefully under it
(we discuss this in Section 5.3). Owner key loss is a known limitation
of any owner-encrypted system and renders past receipts undecryptable;
key escrow is out of scope for v0.1.

We also assume at least one service the agent calls is honest and
uncompromised within the relevant time window. If every service colludes
with the agent operator, Sello cannot detect what they jointly
fabricate. The system's effectiveness scales with the plurality of
independent services.

\hypertarget{design-goals}{%
\subsubsection{3.3 Design goals}\label{design-goals}}

A protocol meeting our threat model must provide the following
properties:

\textbf{P1. Receiver-side signing.} The signing key is held by the
service receiving the agent's call. Not the agent, not the operator, not
any gateway or proxy in the operator's trust boundary. This is the
inversion that makes the rest possible: it places the signer in the only
position that observes the call directly while being independent of the
parties with incentive to misrepresent it.

\textbf{P2. Encryption to owner.} Receipt contents are encrypted to the
agent owner's public key using asymmetric authenticated encryption. The
owner's public key must be bound to the authorization token by a
mechanism that cannot be subverted by an adversary substituting a
different recipient. This property is what makes it safe to publish
receipts on shared public infrastructure.

\textbf{P3. Public transparency log.} Receipts are appended to an
append-only Merkle log with witness cosigning. The log provides
tamper-evidence for logged receipts and global verifiability of
inclusion. Completeness of retrieval is a separate property and depends
on the query mechanism (Section 6).

\textbf{P4. Owner-side discovery.} The owner queries the log by
authorization-token-derived reference, decrypts receipts locally with
their private key, and verifies signatures against a registry of service
identities. No party in the middle reconstructs the trail; no party in
the middle can see it.

A protocol that satisfies P1 alone is ``receiver-attested'' but its
receipts cannot safely live on public infrastructure (privacy leak). A
protocol that satisfies P1+P2 can publish to a private log but inherits
the trust assumptions of that log's operator. A protocol that satisfies
P1+P2+P3 still requires P4 to be useful to owners who don't control the
infrastructure agents run on. The four properties together are what
makes the architecture work.

\begin{center}\rule{0.5\linewidth}{0.5pt}\end{center}

\hypertarget{the-sello-protocol}{%
\subsection{4 The Sello Protocol}\label{the-sello-protocol}}

\hypertarget{receipt-structure}{%
\subsubsection{4.1 Receipt structure}\label{receipt-structure}}

A Sello receipt is a COSE\_Sign1 envelope wrapping an HPKE-encrypted
payload. The structure is:

\[\text{receipt} = \text{COSE\_Sign1}(\text{HPKE.Seal}_{pk_{\text{owner}}}(\text{body}), \text{sk}_{\text{service}})\]

Where:

\begin{itemize}
\tightlist
\item
  \(pk_{\text{owner}}\) is the owner's X25519 public key, bound to the
  authorization token (Section 4.2)
\item
  \(sk_{\text{service}}\) is the service's Ed25519 private signing key
\item
  \(\text{body}\) is a CBOR-encoded structure describing the action
\end{itemize}

The body is defined in CDDL \citep{RFC8610} notation, where \texttt{?}
denotes an optional field:

\begin{verbatim}
receipt-body = {
  agent-identifier:        tstr,     ; derived from token hash
  action-type:             tstr,     ; e.g. "tools/call"
  action-input-hash:       bstr,     ; SHA-256 over canonicalized input
  action-output-hash:      bstr,     ; SHA-256 over canonicalized output
  result-status:           tstr,     ; "success" | "error" | "denied"
  timestamp:               tdate,    ; RFC 3339 UTC timestamp
  ? service-defined-fields: any
}
\end{verbatim}

Action inputs and outputs are referenced by SHA-256 hash, not included
verbatim, to keep receipts small and to limit what the owner can
reconstruct without independent access to the inputs.

The COSE\_Sign1 protected header carries:

\begin{itemize}
\tightlist
\item
  \texttt{alg\ =\ -8} (EdDSA / Ed25519)
\item
  \texttt{kid} (binary key identifier for the service's signing key)
\item
  \texttt{sello\_version\ =\ "0.1.0"}
\item
  \texttt{sello\_token\_ref\ =\ SHA-256(authorization-token-bytes)}
\item
  \texttt{sello\_log\_url} (canonical URL of the log this receipt was
  published to)
\end{itemize}

The protected header is what the receiver uses to look up the service's
public key and the log's identity at verification time.

\hypertarget{the-jws-owner-key-binding}{%
\subsubsection{4.2 The JWS owner-key
binding}\label{the-jws-owner-key-binding}}

The central novel mechanism in Sello is how the receiving service learns
which public key to encrypt to. The owner cannot publish a static
directory of public keys (the agent could call any service, and most
services have no way to look up ``who is this agent's owner''). The
agent cannot present the owner's pubkey directly without enabling
substitution attacks.

We solve this by carrying the owner's HPKE public key as a signed claim
inside the agent's authorization token. The v0.1 token profile uses
compact-serialized JWS \citep{RFC7515} whose payload is a UTF-8-encoded
JSON object containing the claim:

\begin{Shaded}
\begin{Highlighting}[]
\FunctionTok{\{}
  \ErrorTok{...}
  \DataTypeTok{"owner\_hpke\_pk"}\FunctionTok{:} \StringTok{"base64url(32{-}byte X25519 public key)"}\FunctionTok{,}
  \DataTypeTok{"sello\_logs"}\FunctionTok{:} \OtherTok{[}\StringTok{"https://log1.example.com"}\OtherTok{,} \ErrorTok{...}\OtherTok{]}
\FunctionTok{\}}
\end{Highlighting}
\end{Shaded}

The JWS signature, verifiable against the token issuer's public key,
cryptographically binds the owner's HPKE public key to the authorization
token. An adversary who substitutes a different pubkey invalidates the
JWS. An adversary who reuses a captured token cannot change the
recipient.

When a service receives an agent call, it:

\begin{enumerate}
\def\labelenumi{\arabic{enumi}.}
\tightlist
\item
  Verifies the JWS signature against the token issuer's verification
  key.
\item
  Extracts \texttt{owner\_hpke\_pk} from the verified claims.
\item
  Uses it as the HPKE recipient key when encrypting the receipt payload.
\end{enumerate}

This mechanism is, as far as we have been able to determine, novel. We
searched OAuth, JWS, and JWT specifications and extensions for prior art
on token-bound encryption-recipient public keys and found none. The
closest existing mechanism is DPoP \citep{RFC9449}, which binds a
\emph{signing} public key to a token for proof-of-possession. Sello's
mechanism binds an \emph{encryption} public key for confidential
receipts; the cryptographic direction is opposite.

\hypertarget{protocol-flow}{%
\subsubsection{4.3 Protocol flow}\label{protocol-flow}}

When an agent calls a service:

\begin{enumerate}
\def\labelenumi{\arabic{enumi}.}
\tightlist
\item
  The service receives the call and verifies the agent's authorization
  token (JWS signature against the issuer's key).
\item
  The service performs the requested action (or refuses, if its own
  authorization policy denies).
\item
  The service constructs a receipt body describing the action.
\item
  The service encrypts the body to \texttt{owner\_hpke\_pk} using HPKE
  single-shot mode.
\item
  The service signs the encrypted envelope using COSE\_Sign1 with its
  Ed25519 private key.
\item
  The service submits the signed envelope to a transparency log listed
  in the token's \texttt{sello\_logs} claim (or by local policy if the
  claim is absent).
\item
  The log returns an inclusion proof.
\end{enumerate}

When the owner wishes to reconstruct the agent's activity:

\begin{enumerate}
\def\labelenumi{\arabic{enumi}.}
\tightlist
\item
  The owner computes \texttt{sello\_token\_ref\ =\ SHA-256(token-bytes)}
  over the same raw token bytes presented by the agent.
\item
  The owner queries each trusted log for entries matching
  \texttt{sello\_token\_ref} in their protected header.
\item
  For each returned envelope, the owner verifies that its
  \texttt{sello\_log\_url} matches the log that returned it.
\item
  The owner verifies the Merkle inclusion proof against the witnessed
  log root.
\item
  The owner resolves the signing service via the identity registry using
  \texttt{kid} from the protected header, obtaining the service's public
  key.
\item
  The owner verifies the COSE\_Sign1 signature against that public key.
\item
  The owner decrypts the HPKE payload using their HPKE private key, with
  the protected header as additional authenticated data.
\item
  The owner inspects the decrypted receipt body.
\end{enumerate}

A receipt is valid only if all of steps 3 through 7 succeed.

\hypertarget{service-identity-and-revocation}{%
\subsubsection{4.4 Service identity and
revocation}\label{service-identity-and-revocation}}

Sello v0.1 specifies a JSON-file identity registry mapping \texttt{kid}
to \texttt{(service\_identifier,\ public\_key)}. The registry is signed
by a trust root operator and served at a stable URL. Owners cache the
registry with a freshness bound of 24 hours.

Rotation is additive: new keys get new \texttt{kid}s; old \texttt{kid}s
remain valid for verifying receipts signed before rotation.

Revocation uses a sibling \texttt{revoked} table mapping each revoked
\texttt{kid} to a \texttt{revoked\_at} timestamp. \textbf{Receipts are
rejected if their \texttt{kid} appears in \texttt{revoked} AND the
transparency log's integrated time for that receipt is at or later than
\texttt{revoked\_at}.} Receipts whose integrated time precedes
\texttt{revoked\_at} remain verifiable, preserving the historical record
for routine key rotation.

The use of the transparency log's \emph{integrated time}, the timestamp
the log itself assigns to the witnessed entry, rather than the service's
self-asserted timestamp in the receipt body is deliberate. A compromised
service can backdate its \texttt{timestamp} field arbitrarily; it cannot
backdate when the log integrated the entry. We propose this as a
separately defensible novelty: tying revocation decisions to
log-integrated time rather than signer-asserted time is, to our
knowledge, not present in current transparency-log deployments or AI
agent receipt protocols.

The JSON-file registry is a v0.1 stopgap, not a production identity
solution. Section 8 discusses paths forward.

\begin{center}\rule{0.5\linewidth}{0.5pt}\end{center}

\hypertarget{security-analysis}{%
\subsection{5 Security Analysis}\label{security-analysis}}

We analyze Sello's properties under the adversary model from Section
3.2. The analysis is informal but tied to specific protocol mechanisms.

\hypertarget{integrity-of-recorded-actions}{%
\subsubsection{5.1 Integrity of recorded
actions}\label{integrity-of-recorded-actions}}

\textbf{Claim:} A receipt signed by a legitimate service's key cannot be
modified without detection.

\textbf{Argument:} The COSE\_Sign1 signature covers the entire envelope
including the protected header and the HPKE-encrypted payload. Any
modification of the protected header, the payload, or the signature
itself invalidates the signature. Verification against the service's
public key (resolved via the registry by \texttt{kid}) will fail. This
property relies on the unforgeability of Ed25519 under the standard
EUF-CMA assumption, which is well-established
\citep{bernstein2012ed25519}.

\hypertarget{confidentiality-of-receipt-contents}{%
\subsubsection{5.2 Confidentiality of receipt
contents}\label{confidentiality-of-receipt-contents}}

\textbf{Claim:} A passive observer of the transparency log cannot read
receipt contents.

\textbf{Argument:} Receipt contents are HPKE-sealed to the owner's
X25519 public key. HPKE single-shot encryption provides IND-CCA2
security under the standard X25519 and ChaCha20-Poly1305 assumptions
\citep{RFC9180}. The encapsulated key and ciphertext together reveal
nothing about the plaintext without possession of the owner's private
key. The protected header is not encrypted and does reveal metadata: the
\texttt{kid} (which service signed), the \texttt{sello\_token\_ref} (the
token-derived identifier the agent presented), and the
\texttt{sello\_log\_url}. We address this metadata leakage in Section 8.

\hypertarget{independence-of-attestation}{%
\subsubsection{5.3 Independence of
attestation}\label{independence-of-attestation}}

\textbf{Claim:} An adversary that compromises the agent or controls the
operator cannot forge receipts.

\textbf{Argument:} Receipt validity requires a signature by a service's
private Ed25519 key. By the adversary model, the adversary does not
possess service signing keys. The adversary can submit arbitrary bytes
to the log, but those bytes will fail signature verification at step 6
of the owner's verification flow and be rejected. Forgery requires the
additional step of compromising at least one service's signing key,
which is a structurally harder attack than compromising the agent.

If a service's key is compromised, the adversary can forge receipts
signed by that key. The integrated-time revocation rule (Section 4.4)
bounds the damage to receipts whose log integrated time falls between
the compromise and the published \texttt{revoked\_at}. Receipts
predating the compromise remain trustworthy; receipts after
\texttt{revoked\_at} are rejected. This is weaker than perfect forward
security but is the best we can do without per-call key rotation, which
has its own costs.

\hypertarget{tamper-evidence-of-the-receipt-set}{%
\subsubsection{5.4 Tamper-evidence of the receipt
set}\label{tamper-evidence-of-the-receipt-set}}

\textbf{Claim:} Removal or modification of a logged receipt is
detectable.

\textbf{Argument:} The Merkle log structure ensures that any change to a
previously logged entry alters the Merkle root. Witness cosigning
ensures the root cannot be silently rewritten by the log operator. The
owner verifies inclusion proofs against the witnessed root.

This is a cryptographic guarantee, but it is a per-receipt guarantee,
not a set-completeness guarantee. Tamper-evidence applies to receipts
that are \emph{queried and returned}; it does not apply to receipts that
the log operator chooses not to return. We treat this distinction as
important enough to discuss separately, in Section 6.

\hypertarget{known-attacks-sello-does-not-prevent}{%
\subsubsection{5.5 Known attacks Sello does not
prevent}\label{known-attacks-sello-does-not-prevent}}

\textbf{Suppression attack.} The most fundamental limitation. If the
agent simply does not call any services, no receipts are produced. The
owner sees an empty log and cannot distinguish ``agent did nothing''
from ``agent did things off-network.'' Missing receipts are at best an
indirect signal of misbehavior. We sketch a possible mitigation in
Section 8 (mandatory heartbeat receipts from a designated anchor set)
but do not solve it in v0.1.

\textbf{Service collusion.} If a service colludes with the agent
operator to emit false receipts, Sello cannot detect this from a single
receipt. The mitigation is plurality: if a real action involved calls to
multiple services, only the genuine action has receipts from all of
them. A receipt purporting to record an action that no other involved
service witnessed is at minimum suspicious. We do not formalize this
mitigation; it is a defense-in-depth, not a cryptographic guarantee.

\textbf{Service replay.} A service holding its own signing key can
re-emit a previously observed event by re-encrypting and re-signing it,
producing a new envelope that passes verification but represents
activity that did not occur. HPKE single-shot uses fresh randomness per
call, so the new envelope is not byte-identical to the original. We
specify an owner-side deduplication rule based on the tuple
\texttt{(kid,\ sello\_token\_ref,\ timestamp-truncated-to-seconds,\ action-type,\ action-input-hash,\ action-output-hash)};
receipts sharing this tuple are treated as a single logical event.
Replays at higher temporal resolution than seconds are not caught by
this rule but are flagged as anomalous.

\textbf{Token reference enumeration.} The \texttt{sello\_token\_ref} is
a deterministic hash of the authorization token. Anyone who holds the
token can compute the same hash and query the log for all receipts
associated with it. Authorization tokens used with Sello must contain at
least 128 bits of unpredictable entropy. We discuss PIR-based
mitigations in Section 8.

\textbf{Owner key loss.} If the owner loses their HPKE private key, all
past receipts become permanently undecryptable. Key escrow and recovery
are out of scope for v0.1.

\begin{center}\rule{0.5\linewidth}{0.5pt}\end{center}

\hypertarget{completeness-and-retrieval}{%
\subsection{6 Completeness and
Retrieval}\label{completeness-and-retrieval}}

The security analysis in Section 5 establishes per-receipt cryptographic
guarantees: a returned receipt is unforgeable, its contents are
confidential, its tamper is detectable. These are necessary properties
for a verifiable record, but they are not sufficient for what an owner
actually wants, which is the \emph{complete set} of receipts associated
with a particular agent action or a particular time window. We treat the
gap between \emph{per-receipt verifiability} and \emph{set-completeness}
as a distinct concern from cryptographic security and discuss it here.

\textbf{The retrieval gap.} An inclusion proof answers ``is this receipt
in the log?'' It does not answer ``did the log return every matching
receipt?'' A log operator (or a metadata indexer in front of the log)
that returns ten receipts when twenty exist passes every cryptographic
check on each of the ten while withholding the other ten. The owner sees
a verifiable set; the owner does not see that the set is incomplete.

This matters for Sello in particular because owner-side discovery uses
\texttt{sello\_token\_ref} as the lookup key (Section 4.3). The owner
asks the log ``give me all receipts with this token reference.'' A
passive observer cannot tell whether the returned set is complete or
whether some receipts have been silently omitted. The receipts that are
returned remain verifiable. The receipts that are missing are missing.

\textbf{Sources of incompleteness.} There are three distinct failure
modes. The first is operator omission: the log operator or metadata
indexer chooses, accidentally or adversarially, not to return a matching
entry. The second is index incompleteness: the log's underlying Merkle
structure contains the entry, but the auxiliary metadata index used to
satisfy queries does not. Sigstore Rekor, for instance, supports queries
against indexed fields but does not index arbitrary COSE
protected-header values, so queries on \texttt{sello\_token\_ref} may
not be natively supported by Rekor v0 without additional infrastructure.
The third is owner-side query failure: the owner queries one log when
receipts exist in another, missing them entirely.

\textbf{What completeness requires.} Production deployments that need
completeness guarantees have three options. First, \emph{authenticated
query results}: the log can sign a response of the form ``these are the
entries matching this query, and the matching set is exhaustive as of
log tree size N.'' The SCITT working group's SCRAPI draft
\citep{scitt-scrapi} points in this direction. Second, \emph{full log
audit}: the owner downloads the full log (or relevant subsets) and scans
it locally, treating the log as a queryable database rather than asking
the operator for filtered results. This is expensive but defeats
operator omission entirely. Third, \emph{multi-log redundancy}: receipts
are submitted to multiple independent logs, and the owner queries all of
them; a single dishonest log cannot suppress a receipt without colluding
with every other log it was submitted to.

\textbf{What Sello v0.1 provides.} The reference implementation includes
a Rekor discovery adapter that explicitly documents itself as
discovery-only, not completeness-proving. The published
\texttt{sello\_log\_url} in the protected header tells the owner
\emph{which} log to query, but does not commit the log to returning
exhaustive results. A future version of Sello may specify an
authenticated query mechanism profiled against SCITT-SCRAPI; we do not
solve it in v0.1.

\textbf{Position summary.} Sello v0.1 provides strong per-receipt
security and discovery-aided retrieval. It does not provide
cryptographic set-completeness. Deployments where set-completeness
matters (audit reconstruction, regulatory evidence) should plan for one
of the three completeness mechanisms above.

\begin{center}\rule{0.5\linewidth}{0.5pt}\end{center}

\hypertarget{evaluation}{%
\subsection{7 Evaluation}\label{evaluation}}

We present first-party microbenchmarks of the cryptographic operations
Sello performs, measured with the reference implementation against a
local mock transparency log. The measurements establish steady-state
per-receipt costs at the service and the owner. Network latency for
submission to a hosted public log is discussed separately in §7.3
because it depends on the specific log operator's deployment rather than
on the Sello protocol itself.

\hypertarget{methodology}{%
\subsubsection{7.1 Methodology}\label{methodology}}

The reference implementation \citep{sello-repo} ships with a benchmark
tool (\texttt{src/cli/bench.ts}) that exercises the full receipt
lifecycle. For each iteration the bench: (1) constructs a verified
compact JWS authorization token, (2) creates a Sello receipt at the
service side (HPKE seal of the CBOR-encoded receipt body to the owner's
X25519 public key, followed by COSE\_Sign1 signing with the service's
Ed25519 private key), (3) submits the signed envelope to an in-process
mock transparency log that returns an inclusion proof, and (4) performs
full owner-side verification (proof verification, registry resolution,
COSE\_Sign1 signature verification, HPKE open, and receipt body
validation). The mock log replaces network submission with deterministic
in-process bookkeeping; this isolates the cryptographic costs from
log-operator-specific network latency. Log submission latency to a
hosted public log is discussed separately in §7.3.

The cryptographic primitives are Node.js bindings to the system OpenSSL:
X25519 key agreement and Ed25519 signing via \texttt{node:crypto},
ChaCha20-Poly1305 AEAD via \texttt{createCipheriv}, HKDF-SHA256 via
\texttt{createHmac}, and SHA-256 via \texttt{createHash}. HPKE is
implemented per RFC 9180 base mode with the suite specified in §2.1; the
implementation is pinned to RFC 9180 Appendix A.2.1 test vectors and
validated in the test suite.

Measurements were taken on an Apple M2 (8 physical cores, arm64), 24 GB
RAM, macOS 26.2 (Darwin kernel 25.2.0), Node v24.16.0 with V8
13.6.233.17 and OpenSSL 3.5.6, plugged into AC power. The repository
state was pinned to commit \texttt{8b0da95} at the time of measurement.
System thermal pressure was ``Nominal'' throughout the run (Apple
Silicon does not expose absolute CPU die temperature through
\texttt{powermetrics}; thermal pressure is the first-party signal). The
machine was a development laptop with 29 days of uptime and load
averages of 6-10 during the run, representative of a working environment
rather than a dedicated benchmark host; production deployments on
lightly loaded hardware should see equal or tighter tails.

The bench prepends a 500-iteration warmup phase before any timed
measurements to amortize JIT compilation and instruction-cache warmup,
and invokes \texttt{global.gc()} once before warmup with the
\texttt{-\/-expose-gc} flag to start each run from a clean
garbage-collection state. We attempted to disable Spotlight indexing for
the measurement window via \texttt{mdutil\ -a\ -i\ off} but this did not
take effect on macOS 26 (likely due to System Integrity Protection
policy); indexing remained enabled throughout. The tail-latency
improvements over an earlier methodology iteration are therefore
attributable to the warmup and GC controls rather than to Spotlight
isolation.

For each iteration count \(N \in \{10, 100, 1000\}\), we executed the
bench multiple times (3 runs at N=10, 5 runs at N=100, 10 runs at
N=1000), with a 3-second sleep between runs to allow thermal effects to
settle. Per-receipt operations (receipt creation, single-receipt
verification) record every individual iteration duration using
\texttt{performance.now()}. From each run we compute the per-run median,
p95, and p99 over the iteration samples. We then take the median across
runs of each statistic to be robust to outliers. Batch verification is
one timed event per run, so we report the median across runs of the
per-run total and the per-run per-receipt average.

\hypertarget{results}{%
\subsubsection{7.2 Results}\label{results}}

The reference implementation produces fixed-size receipts independent of
action input or output (which are referenced by SHA-256 hash, not
included verbatim). Table 1 reports the wire-format sizes for the v0.1
envelope.

\textbf{Table 1: Wire-format sizes.}

\begin{longtable}[]{@{}ll@{}}
\toprule\noalign{}
Component & Size \\
\midrule\noalign{}
\endhead
\bottomrule\noalign{}
\endlastfoot
CBOR-encoded receipt body & 234 bytes \\
COSE\_Sign1 protected header & 112 bytes \\
HPKE payload (enc \textbar\textbar{} ciphertext) & 282 bytes \\
Full COSE\_Sign1 envelope & 467 bytes \\
Mock log proof (JSON) & 253 bytes \\
\end{longtable}

A complete Sello envelope on the wire is just under half a kilobyte;
this is independent of action size because action inputs and outputs are
referenced only by their SHA-256 digests inside the receipt body.

Table 2 reports per-receipt cryptographic latency at three iteration
counts. Each cell is the median across runs of the corresponding per-run
statistic.

\textbf{Table 2: Per-receipt cryptographic latency (milliseconds).}

\begin{longtable}[]{@{}
  >{\raggedright\arraybackslash}p{(\columnwidth - 8\tabcolsep) * \real{0.2000}}
  >{\raggedright\arraybackslash}p{(\columnwidth - 8\tabcolsep) * \real{0.2000}}
  >{\raggedright\arraybackslash}p{(\columnwidth - 8\tabcolsep) * \real{0.2000}}
  >{\raggedright\arraybackslash}p{(\columnwidth - 8\tabcolsep) * \real{0.2000}}
  >{\raggedright\arraybackslash}p{(\columnwidth - 8\tabcolsep) * \real{0.2000}}@{}}
\toprule\noalign{}
\begin{minipage}[b]{\linewidth}\raggedright
Operation
\end{minipage} & \begin{minipage}[b]{\linewidth}\raggedright
N
\end{minipage} & \begin{minipage}[b]{\linewidth}\raggedright
median (p50)
\end{minipage} & \begin{minipage}[b]{\linewidth}\raggedright
p95
\end{minipage} & \begin{minipage}[b]{\linewidth}\raggedright
p99
\end{minipage} \\
\midrule\noalign{}
\endhead
\bottomrule\noalign{}
\endlastfoot
Receipt creation (service) & 10 & 0.496 & 0.741 & 0.770 \\
Receipt creation (service) & 100 & 0.471 & 1.016 & 2.243 \\
Receipt creation (service) & 1000 & \textbf{0.454} & \textbf{0.954} &
\textbf{1.534} \\
Verification, single (owner) & 10 & 0.393 & 1.072 & 1.245 \\
Verification, single (owner) & 100 & 0.305 & 0.656 & 0.854 \\
Verification, single (owner) & 1000 & \textbf{0.281} & \textbf{0.573} &
\textbf{1.034} \\
Verification, batch per receipt & 10 & 0.316 & n/a & n/a \\
Verification, batch per receipt & 100 & 0.348 & n/a & n/a \\
Verification, batch per receipt & 1000 & \textbf{0.320} & n/a & n/a \\
\end{longtable}

We treat the N=1000 row as steady-state and report it as the headline
result: a Sello-aware service adds \textbf{approximately 0.45 ms of
median CPU work per agent call} (HPKE seal of the receipt body,
COSE\_Sign1 signing, in-process log append), and an owner reconstructing
activity does \textbf{approximately 0.28 ms of median CPU work per
receipt} (Merkle proof verify, registry lookup, COSE\_Sign1 verify, HPKE
open, receipt body decode and validate). A service handling 1000 calls
per second per process would spend approximately 0.45 seconds of CPU per
second on Sello receipt creation. An owner reconstructing a day of
activity from 100,000 receipts can do so in approximately 32 seconds of
CPU time using batch verification.

\textbf{Variance and tail behavior.} Across runs, the per-run median is
tight: at N=1000, receipt creation medians range from 0.449 to 0.479 ms
across ten runs (a 6.7\% spread), and single-receipt verification
medians range from 0.275 to 0.292 ms (a 6.2\% spread). Tail latencies
are also stable: across the ten N=1000 runs, all ten verify-one-receipt
p99 values fell within 2x of the median p99, and nine of ten
create-receipt p99 values fell within 2x of the median p99. The widest
p99 observed for any single run was 3.073 ms for receipt creation and
1.487 ms for verification. Production deployments on dedicated hardware
should see equal or tighter tails; the measurements here were taken on a
working laptop with load averages of 6-10 during the run.

\textbf{Warmup effect.} Even with a 500-iteration warmup, a residual
warmup effect remains visible at low iteration counts: single-receipt
verification drops from 0.393 ms median at N=10 to 0.281 ms at N=1000, a
29\% reduction. This is consistent with V8 tiered compilation continuing
to optimize hot code paths beyond the warmup window. The N=1000 numbers
should be treated as steady-state; the N=10 numbers approximate the
cold-start cost a process pays for its first few receipts.

\hypertarget{log-submission-latency-not-measured-first-party}{%
\subsubsection{7.3 Log submission latency (not measured
first-party)}\label{log-submission-latency-not-measured-first-party}}

Submission of a signed envelope to a hosted public transparency log adds
a network round-trip and, for batched log designs like Sigstore Rekor
v2, additional time for batch integration. The v0.1 reference
implementation does not measure log submission latency first-party
because it depends on the specific log operator's deployment
characteristics rather than on the Sello protocol itself. In a
production deployment, services should treat log submission as an
asynchronous operation: the service can return the action response to
the agent immediately and submit the receipt to the log in the
background, optionally including the eventual inclusion proof in a
subsequent receipt or an out-of-band channel. The protocol flow (§4.3)
is compatible with this: the log-submission step is not on the critical
path of the agent call.

\hypertarget{comparison-with-adjacent-systems}{%
\subsubsection{7.4 Comparison with adjacent
systems}\label{comparison-with-adjacent-systems}}

We compare Sello against the seven adjacent receipt protocols surveyed
in §2.2, scored against the four design properties P1-P4.

\textbf{Table 3: Property comparison against adjacent receipt
protocols.}

\begin{longtable}[]{@{}lllll@{}}
\toprule\noalign{}
System & P1 & P2 & P3 & P4 \\
\midrule\noalign{}
\endhead
\bottomrule\noalign{}
\endlastfoot
Agent Receipts & no & no & no & no \\
Signet (v0.10) & no & partial & no & no \\
Pipelock & no & no & no & no \\
Agent Passport System & no & no & no & no \\
draft-farley-acta & no & no & no & no \\
draft-nivalto-agentroa & no & no & yes & no \\
Attested Intelligence & no & no & partial & no \\
\textbf{Sello} & yes & yes & yes & yes \\
\end{longtable}

Notes: ``no'' indicates the property is not present. ``partial''
indicates a partial form of the property is present but does not meet
the full design goal as specified in §3.3. Signet's encryption is
symmetric and reuses the signing identity, which is not the same as P2's
asymmetric encryption to an independent owner key. Attested Intelligence
ships tamper-evident logging but the logs are point-to-point evidence
bundles rather than public Merkle logs.

\begin{center}\rule{0.5\linewidth}{0.5pt}\end{center}

\hypertarget{discussion}{%
\subsection{8 Discussion}\label{discussion}}

\hypertarget{limitations-sello-does-not-address-in-v0.1}{%
\subsubsection{8.1 Limitations Sello does not address in
v0.1}\label{limitations-sello-does-not-address-in-v0.1}}

Three known limitations bear emphasizing because they shape what Sello
is and isn't useful for.

\textbf{Adoption-incentive problem.} The most fundamental limitation.
Sello requires services to choose to emit receipts. Services have no
inherent incentive to do so, especially in v0.1 before any regulatory
mandate exists. We discuss possible incentive structures in Section 8.3
but do not solve them. Sello is only useful if services adopt it; in
2026, that adoption is zero.

\textbf{Suppression attack.} Discussed in Section 5.5. The agent simply
not calling services produces no receipts. We sketch a
mandatory-heartbeat mitigation in Section 8.3 but do not specify it in
v0.1.

\textbf{Server identity bootstrapping.} The JSON-file registry signed by
a trust root is a placeholder. Production deployments need either
Sigstore-Fulcio-style OIDC keyless identities, DNS-based key publication
(DNSSEC or DNS-over-HTTPS with key records), or a Certificate
Transparency-style PKI for services. We expect convergence on a
Sigstore-Fulcio approach but defer the choice to v0.2.

\hypertarget{position-relative-to-existing-observability}{%
\subsubsection{8.2 Position relative to existing
observability}\label{position-relative-to-existing-observability}}

Sello is not a replacement for existing agent observability tools.
Langfuse, LangSmith, and their peers tell the owner what the agent
\emph{thought} it was doing: the prompts, the reasoning traces, the tool
selection logic. This is essential information for debugging agent
behavior and understanding why an agent did what it did.

Sello tells the owner what the world \emph{saw} the agent do. The two
views are complementary. Mature deployments will want both. Sello
receipts can serve as ground truth against which agent self-reports are
checked; divergence between self-report and witnessed receipts is itself
a high-signal anomaly. We discuss this divergence-detection use case in
Section 8.3 as a future extension.

\hypertarget{future-extensions}{%
\subsubsection{8.3 Future extensions}\label{future-extensions}}

We sketch four extensions that build on the v0.1 protocol. None are
implemented; we present them as future-work directions and as
illustrations of the design space the protocol opens.

\textbf{Regulator-as-second-recipient.} HPKE supports multi-recipient
sealing via COSE\_Encrypt. A receipt could be sealed to two public keys:
the owner's, and a regulator's published HPKE pubkey listed in a
jurisdiction-bound registry (e.g., an EU AI Act Article 26
deployer-registry entry). The regulator could decrypt during an audit
without the owner's cooperation, but only with their specific key. This
would make Sello a natural compliance primitive for high-risk AI systems
under the EU AI Act \citep{euaiact}. We have not seen this mechanism in
any existing protocol.

\textbf{Suppression detection via mandatory heartbeat receipts.} A
subset of services designated as an ``anchor set'' in the agent's
authorization token could be required to emit heartbeat receipts every
\(N\) seconds whether or not the agent called them. Missing heartbeats
during a session window would constitute positive evidence of
suppression. The hard part is making heartbeats themselves
non-suppressible (the operator can drop heartbeat traffic); we propose
layered direct-to-owner heartbeats as a partial fix. We are not aware of
any agent protocol that addresses suppression cryptographically.

\textbf{Inverse observability: divergence detection.} Combine Sello
receipts with agent-emitted observability traces (Langfuse, LangSmith,
etc.) and detect divergence between the two. If the agent's
self-reported trace claims a call that no receipt witnesses, either the
agent is lying or the service is colluding. The Verifiability-First
paper \citep{gupta-verifiabilityfirst} proposes a related concept using
a secondary ``Audit Agent'' but does not ground it in cryptographic
witnessing. Sello receipts could provide exactly that grounding.

\textbf{Private information retrieval.} The \texttt{sello\_token\_ref}
enumeration attack (Section 5.5) could be mitigated by Private
Information Retrieval over the log. PIR over Certificate Transparency
logs has been studied \citep{kales-pir-ct}; applying the same techniques
to Sello is straightforward but adds significant cost. A PIR mode in
v0.2 would be appropriate for high-privacy deployments.

\hypertarget{the-adoption-incentive-question}{%
\subsubsection{8.4 The adoption-incentive
question}\label{the-adoption-incentive-question}}

We close with the most honest question. Why would any service emit Sello
receipts? In 2026, the answer is ``they wouldn't.'' There is no
regulatory mandate, no commercial pressure, no end-user demand.

Three incentive structures could change this:

\begin{enumerate}
\def\labelenumi{\arabic{enumi}.}
\item
  \textbf{Regulatory mandate.} EU AI Act Article 12 requires automatic
  logging for high-risk AI systems; Article 26 places six-month
  retention obligations on deployers. Sello is a natural primitive for
  satisfying these obligations cryptographically. If the EU Commission
  or a similar body specifies a receipt format, Sello (or something like
  it) becomes mandatory and adoption follows.
\item
  \textbf{Agent-commerce settlement coupling.} Agent payment protocols
  (Stripe Machine Payments, Skyfire KYAPay, the Cloudflare/Stripe agent
  commerce protocol) are emerging in 2026. If receipt emission is
  coupled to settlement, so that services that emit receipts settle and
  services that don't, don't, adoption is automatic. This is a market
  design problem more than a protocol design problem.
\item
  \textbf{Trust premium.} Services that emit receipts may command higher
  trust and therefore higher prices in agent-mediated marketplaces. This
  requires owners to value receipts enough to pay for them, which
  requires educated users; bootstrapping is slow but possible.
\end{enumerate}

We do not solve the incentive problem in this paper. We note that all
three paths require Sello (or an equivalent protocol) to exist first.
The protocol is the precondition; the incentive structure is the next
problem.

\begin{center}\rule{0.5\linewidth}{0.5pt}\end{center}

\hypertarget{conclusion}{%
\subsection{9 Conclusion}\label{conclusion}}

AI agent observability is structurally compromised when the agent is the
source of truth for its own behavior. The fix is the same architectural
inversion humans have used since 3000 BCE: the party recording an event
should be independent of the party being recorded.

Sello applies this inversion to AI agent actions through four combined
properties: receiver-side signing by the called service, HPKE encryption
to an owner public key bound to the authorization token, publication to
a witness-cosigned Merkle log, and owner-side discovery by token
reference. The combination is novel to our knowledge; no surveyed system
combines all four. The cryptographic primitives are standard
(COSE\_Sign1, HPKE, Ed25519, transparency logs). The contribution is the
architectural composition and the JWS owner-key binding that makes it
work.

We have been explicit about what Sello does not solve: the suppression
attack, service collusion, the adoption-incentive problem. The sketched
extensions (regulator-recipient mode, heartbeat-based suppression
detection, inverse observability for trace verification) are not
required for v0.1 to be useful, and we have not built them.

The protocol and reference implementation are published at
https://sello.build under CC BY 4.0. Review, adversarial testing, and
pull requests are welcome. Where alignment is natural with adjacent
projects (Signet, the SCITT working group, AgentROA), we'll seek
collaboration. The architecture matters more than which project ships it
first; what matters is that AI agent observability stops trusting the
agent to be its own witness.

The agent is not the right narrator of its own actions. The services it
touches are.

\begin{center}\rule{0.5\linewidth}{0.5pt}\end{center}

\hypertarget{acknowledgments}{%
\subsection{Acknowledgments}\label{acknowledgments}}

This work benefited from public writing by William Hou (Signet), Tymofii
Pidlisnyi (APS), Otto Jongerius (Agent Receipts), Joseph Michalak
(AgentROA), and Jing Zhang (Right to History). Any errors are mine.

\begin{center}\rule{0.5\linewidth}{0.5pt}\end{center}

  \bibliography{refs.bib}

\end{document}